# Pattern Informatics and Its Application for Optimal Forecasting of Large Earthquakes in Japan

K. Z. NANJO[1,2], J. B. RUNDLE[2], J. R. HOLLIDAY[2], and D. L. TURCOTTE[3]

[1]The Institute of Statistical Mathematics, Minato-ku, Tokyo 106-8569, Japan

[2]Center for Computational Science and Engineering, c/o Department of Physics, University of California at Davis, One Shields Avenue, Davis, CA 95616, USA

[3]Department of Geology, University of California at Davis, One Shields Avenue, Davis, CA 95616, USA

E-mail: nanjo@cse.ucdavis.edu; jbrundle@ucdavis.edu; holliday@physics.ucdavis.edu; turcotte@geology.ucdavis.edu

Corresponding author

K. Z. NANJO

Center for Computational Science and Engineering, c/o Department of Physics, University of California at Davis, One Shields Avenue, Davis, CA 95616, USA

E-mail: nanjo@cse.ucdavis.edu

Tel: +1-530-752-6419

Fax: +1-530-754-4885

Abbreviated title: Pattern Informatics for Earthquake Forecast

*Abstract*-Pattern informatics (PI) technique can be used to detect precursory seismic activation or quiescence and make earthquake forecast. Here we apply the PI method for optimal forecasting of large earthquakes in Japan, using the data catalogue maintained by the Japan Meteorological Agency. The PI method is tested to forecast large (magnitude $m \geq 5$) earthquakes for the time period 1995-2004 in the Kobe region. Visual inspection and statistical testing show that the optimized PI method has forecasting skill, relative to the seismic intensity data often used as a standard null hypothesis. Moreover, we find a retrospective forecast that the 1995 Kobe earthquake ($m = 7.2$) falls in a seismically anomalous area. Another approach to test the forecasting algorithm is to create a future potential map for large ($m \geq 5$) earthquake events. This is illustrated using the Kobe and Tokyo regions for the forecast period 2000-2009. Based on the resulting Kobe map we point out several forecasted areas: the epicentral area of the 1995 Kobe earthquake, the Wakayama area, the Mie area, and the Aichi area. The Tokyo forecasted map was created prior to the occurrence of the Oct. 23, 2004 Niigata earthquake ($m = 6.8$) and the principal aftershocks with $5.0 \leq m$. We find that these events occurred in a forecasted area in the Tokyo map. The PI technique for regional seismicity observation substantiates an example showing considerable promise as an intermediate-term earthquake forecasting in Japan.

**Key words**: Pattern informatics, earthquake, forecasting, seismicity, 1995 Kobe earthquake, 2004 Niigata earthquake.

1. *Introduction*

Earthquakes have great scientific, societal, and economic significance. The 17 Jan. 1995 Kobe, Japan, earthquake (hereinafter referred as the 1995 Kobe earthquake) was only a magnitude $m = 7.2$ event and yet killed nearly 6,000 persons and produced an estimated $200 billion loss. Similar scenarios are possible at any time in San Francisco, Seattle, and other U.S. urban centers along the Pacific plate boundary. The magnitude of potential loss of life and property is so great that reliable earthquake forecasting should be at the forefront of research goals, especially in Japan.

Millions of dollars and thousands of work years have been spent on observational programs searching for reliable precursory phenomena. Possible precursory phenomena include changes in seismicity, changes in seismic velocity, tilt and strain precursors, electromagnetic signals, hydrologic phenomena, and chemical emissions (TURCOTTE, 1991; SCHOLZ, 2002). For example, TSUNOGAI and WAKITA (1995, 1996) found that the ion concentrations of ground water issuing from deep wells located near the epicenter of the 1995 Kobe earthquake showed precursory phenomena (See also JOHNANSEN *et al.*, 1996). A few successes have been reported, but, to date, no precursors to large earthquake have been detected that would provide reliable forecasts (Nature Debates, Denoted on earthquake forecasting, http://www.nature.com/nature/debates/earthquake/, 1999).

The Earth's crust is clearly extremely complex and it is generally accepted that earthquakes are a chaotic phenomenon. Thus, as in the case of weather forecasting, earthquake forecasting must be considered on a statistical basis (RUNDLE *et al.*, 2003). A fundamental question is whether the statistical properties of seismicity patterns can be used to forecast future earthquakes. Premonitory seismicity patterns were found for some strong earthquakes in California and Nevada using algorithm "CN" and for $m > 8$ worldwide using algorithm "M8" (e.g., KEILIS-BOROK, 1990; KEILIS-BOROK and ROTWAIN, 1990; KEILIS-BOROK and KOSSOBOKOV, 1990; KEILIS-BOROK and SOLOVIEV, 2003).

Alternatively, a new approach to earthquake forecasting, the pattern informatics (PI) approach, has been proposed by RUNDLE *et al.* (2002), TIAMPO *et al.* (2002a, b, c), and HOLLIDAY *et al.* (2004a, b).  This approach is based on the strong space-time correlations that are responsible for the cooperative behavior of driven threshold systems and arises both from threshold dynamics as well as from the mean field (long range) nature of the interactions.  The PI technique can be used to detect precursory seismic activation or quiescence and make earthquake forecasts.  Applications to earthquake data from southern California show that the PI method is a powerful technique for forecasting large events, but no one has attempted to apply this method to Japanese earthquakes.  Moreover, it will be interesting to test whether the 1995 Kobe earthquake could have been forecasted using the method.

The purpose of this paper is to study the applicability of the pattern informatics (PI) algorithm for forecasting large earthquakes in Japan.  As an example, we will present a forecast of large ($m \geq 5$) earthquakes during the time period 1995-2004 in the Kobe region: the region that includes the epicenter of the 1995 Kobe earthquake.  First, we will briefly introduce the PI method.  Next, we will describe the earthquake catalogues used in this paper.  Carrying out visual inspections and statistical testing, we will find that the method can has more precision in forecasting large earthquakes than a simple look at where earthquakes have occurred in the past. Then, we will find a retrospective forecast that the 1995 Kobe earthquake fall into a seismically anomalous area.

The true test of the forecasting algorithm is to make a future forecast that proves correct. TIAMPO *et al.* (2002b), RUNDLE *et al.* (2002, 2003), and HOLLIDAY *et al.* (2004a) did this for the period of 2000-2009 in Southern California. In this paper we will also make a future forecast in the time period from 2000 to 2009 for two regions: the Kobe region and the Tokyo region in which the Tokyo metropolitan area is included. This paper will show considerable promise as an intermediate-term event forecasting tool for Japanese earthquakes.

## 2. *Pattern Informatics*

The PI approach is a six step process that creates a time-dependent system state vector in a real valued Hilbert space and uses the phase angle to predict future states (RUNDLE *et* *al.*, 2003). The method is based on the idea that the future time evolution of sesmicity can be described by pure phase dynamics (MORI and KURAMOTO, 1998; RUNDLE *et al.*, 2000a, b). This PI method was originally developed by TIAMPO *et al.* (2002b). HOLLIDAY *et al.* (2004a) showed that the modified PI method recently proposed by KLEIN (2004) has better forecasting skills than the original method. Here we use the modified PI method for forecasting large earthquakes in Japan.

First the study area is divided into $N$ square boxes. The center of the $i$-th box is denoted by $\bar{x}_i$ and each box has an edge length $\Delta x$. The seismic intensity in box $i$ is defined to be the total number of earthquakes $n(\bar{x}_i, t_b, t)$ in the box during the period from the base time $t_b$ to time $t$ ($> t_b$) with magnitude greater than $m_c$. For each box an activity rate function $s(\bar{x}_i, t_b, t)$ is defined to be the average rate of the occurrence of earthquakes in box $i$ during the period $t_b$ to $t$. That is,

$$s(\bar{x}_i, t_b, t) = \frac{n(\bar{x}_i, t_b, t)}{t - t_b}. \tag{1}$$

Values of the average rate of occurrence of earthquakes in box $i$ are obtained by taking the $t_b$ values at daily intervals from $t_0$ to $t$-1. These values are averaged to give its mean value

$$s^*(\bar{x}_i, t_0, t) = \frac{1}{t-1-t_0} \sum_{t_b=t_0}^{t-1} s(\bar{x}_i, t_b, t). \tag{2}$$

The normalized mean value $\hat{s}^*(\bar{x}_i, t_b, t)$ is found by subtracting the spatial mean for all boxes and dividing by the spatial standard deviation

$$\hat{s}^*(\bar{x}_i, t_b, t) = \frac{s^*(\bar{x}_i, t_0, t) - \frac{1}{N}\sum_{j=1}^{N} s^*(\bar{x}_j, t_0, t)}{\sqrt{\frac{1}{N}\sum_{j=1}^{N}\left\{s^*(\bar{x}_j, t_0, t) - \frac{1}{N}\sum_{k=1}^{N} s^*(\bar{x}_k, t_0, t)\right\}^2}}. \tag{3}$$

The change in the normalized mean value is found by subtracting the normalized mean value for the time period $t_0$ to $t_1$ ($> t_0$) from the normalized mean value for the time period $t_0$ to $t_2$ ($> t_1$)

$$\Delta \hat{s}^*(\bar{x}_i, t_0, t_1, t_2) = \hat{s}^*(\bar{x}_i, t_0, t_2) - \hat{s}^*(\bar{x}_i, t_0, t_1). \tag{4}$$

Finally we introduce a probability of change of activity $P(\bar{x}_i, t_0, t_1, t_2)$ in a box. This is related to the square of $\Delta \hat{s}^*(\bar{x}_i, t_0, t_1, t_2)$ by

$$P(\bar{x}_i, t_0, t_1, t_2) = \{\Delta \hat{s}^*(\bar{x}_i, t_0, t_1, t_2)\}^2. \tag{5}$$

Because the $\Delta \hat{s}^*(\bar{x}_i, t_0, t_1, t_2)$ is squared, the probability is a measure of both seismic activation and seismic quiescence.

Schematically, this whole process can be represented by

$$n \to s \to s^* \to \hat{s}^* \to \Delta \hat{s}^* \to P, \tag{6}$$

where the symbol ^ means "apply normalization", the symbol $\Delta$ means "calculate the change in rate", and the symbol * means "average over base times".

We are interested in seismic activation and seismic quiescence relative to the background; the new probability function $P'(\bar{x}_i, t_0, t_1, t_2)$ is defined by the difference between $P(\bar{x}_i, t_0, t_1, t_2)$ and its spatial mean

$$P'(\bar{x}_i, t_0, t_1, t_2) = P(\bar{x}_i, t_0, t_1, t_2) - \frac{1}{N}\sum_{j=1}^{N} P(\bar{x}_j, t_0, t_1, t_2). \tag{7}$$

The use of $P'(\bar{x}_i, t_0, t_1, t_2)$ to forecast earthquakes is referred to as pattern informatics. Forecasts should convey information for time $t$ in the range: $t_3 > t > t_2$. We call the time interval $t_2 - t_1$ the "change interval" and the interval $t_3 - t_2$ the "forecast interval". In this paper, to improve performance, the PI method is optimized by adjusting the length of the change interval ($t_2 - t_1$) and the initial time $t_0$.

### 3. Application of PI Method

We use a seismic catalog maintained by the Japan Meteorological Agency (JMA).

This catalog includes the data of earthquakes with $m \geq 0$ for the time period since 1923 in and around Japan. The relevant data consists of time, magnitude, and location given by east longitude, north latitude, and depth. To ensure the completeness of the earthquake catalog, events of magnitude equal to or larger than the lower cut-off magnitude $m_c = 3$ with depth shallower than 20 km in the time period from Jan. 1, 1955 to the present (as of Mar. 14, 2004) are selected.

To apply the PI method to Japanese earthquakes we need to determine $\Delta x$. Following TIAMPO *et al.* (2002b) we use $\Delta x = 0.1° \approx 11$ km. Boxes of this size correspond roughly to the linear scale size of $m \approx 5 - 6$ earthquakes. We try to forecast earthquakes of $m \geq 5$. The idea is to use information on small events having spatial scales $\lambda < \Delta x$ to forecast the occurrence of large events having scales $\lambda > \Delta x$.

We are also interested in a retrospective forecast of the 1995 Kobe earthquake ($m = 7.2$). The epicenter was 135.03° east longitude and 34.58° north latitude with the depth of 16 km. This earthquake produced a surface rupture with a length of about 9 km appearing along the pre-existing right lateral Nojima fault. Six aftershocks having magnitude $m \geq 5$ followed the main shock along this fault.

To examine whether our forecasting capability is changed if the size of our study regions is changed, we consider 5 study regions: the extent of these regions is changed but they have the same center which is near the epicenter of the 1995 Kobe earthquake. Table 1 summarizes the study regions. The regions 4 and 5 include the epicenter of the Oct. 6, 2000 Tottori earthquake of $m = 7.3$ (hereinafter referred to as the 2000 Tottori earthquake). The epicentre was 133.35° east longitude and 35.27° north latitude with a depth of 9 km.

Before discussing PI method forecast for Kobe region during the period 1995-2004 we describe earthquake data from the Tokyo region (No. 6 in Table 1). The analyzed earthquake data is from the region between 136.0-142.0° east longitude and between 33.0-38.0° north latitude (depth shallower than 20 km). The Tokyo metropolitan area is located at the center of region 6. Eighty-two earthquakes having $m \geq 5$ occurred for the period Jan. 1, 2000-the present (Mar. 14, 2004). An earthquake swarm associated with Miyake volcano started on Jun. 26, 2000 (hereinafter referred to as the 2000 Miyake earthquake swarm). Since then there have been 74 earthquakes with $5.0 \leq m < 6.0$ and six earthquakes with $6.0 \leq m$. This swarm is located at about 33.8-34.3° north latitude and 139.0-139.5° east longitude.

## 4. Forecasting Large Earthquakes in the Kobe Region during the Period 1995-2004

*4.1 Visual Inspection*

The PI approach to forecasting large earthquakes in the Kobe region during the period Jan. 1, 1995- Mar. 14, 2004 is best illustrated using a specific example. A PI-method optimal forecast of earthquake occurrence in region 4 is given in Figure 1. In applying the method to the $N = 1500$ $0.1° \times 0.1° \times 20$ km boxes with $m_c = 3$, the times used are $t_0 =$ Jan. 1, 1960, $t_1 =$ Jan. 1, 1968, $t_2 =$ Dec. 31, 1994, and $t_3 =$ Mar. 14, 2004. The $t_2$-value is just before the occurrence of the 1995 Kobe earthquake. In region 4, there are 18 earthquakes having $m \geq 5$ occurring in the period 1995-2004.

Relative values of the probability of activity are given in the form

$\log_{10}\{P'(\bar{x}_i,t_0,t_1,t_2)/P'_{MAX}\}$ where $P'_{MAX}$ is the maximum in the probabilities $P'(\bar{x}_i,t_0,t_1,t_2)$. The colour coded anomalies are shown in Figure 1. Note that only positive values of $\log_{10}\{P'(\bar{x}_i,t_0,t_1,t_2)/P'_{MAX}\}$ are given. Thus the color-coded regions represent regions of anomalously high seismic activation or high seismic quiescence. The colour-coded anomalies are associated with large ($m \geq 5$) earthquakes for both current (triangles, $t_1 < t < t_2$) and future (circles, $t_2 < t < t_3$) time periods in this figure. Fifteen future earthquakes out of 18 occur either on areas of forecasted anomalous activity or within the margin of error of 11 km (the coarse grained box size). Note that the 1995 Kobe earthquake (123.0° East long., 34.6° North lat.) and the 2000 Tottori earthquake (133.4° East long., 35.3° North lat.) fall into warmer coloured anomalies.

Figure 2 shows seismic intensity $n(\bar{x}_i,t_0,t_2)/n_{MAX}$ using the data from $t_0$ to $t_2$. The large ($m \geq 5$) earthquakes in the change interval (triangles, $t_1 < t < t_2$) and forecast interval (circles, $t_2 < t < t_3$) are included in this figure. Comparing Figures 1 and 2 shows that the optimized PI method narrows the possible locations where large earthquakes are expected.

We also carry out visual inspections for the other regions (Nos. 1, 2, 3, and 5 in Table 1).

We take $t_0$ = Jan. 1, 1960, $t_1$ = Jan. 1, 1968, $t_2$ = Dec. 31, 1994, and $t_3$ = Mar. 14, 2004 as done for region 4.  We find that the 1995 Kobe earthquake again falls into one of the colour coded areas.  Moreover, for region 5 the 2000 Tottori earthquake also falls in a colour coded area.  These visual inspections substantiate that the 1995 Kobe earthquake falls in the forecast area

*4.2 Statistical Testing*

Visual inspection of Figure 1 shows that the retrospective forecast is reasonably successful, but rigorous statistical testing is needed.  For a null hypothesis, we use the actual seismic intensity data in the time period from $t_0$ to $t_2$ $\{n(\bar{x}_i, t_0, t_2)/n_{MAX}\}$ as a probability density, where $n_{MAX}$ is the largest value of $n(\bar{x}_i, t_0, t_2)$.  The use of seismic intensity data has been proposed for the standard null hypothesis (KAGAN and JACKSON, 2000).  This hypothesis was used for testing the PI forecast for southern California earthquakes (TIAMPO *et al.*, 2002a).

To test the PI method forecast we utilize the maximum likelihood test. This is accepted as the standard approach of testing earthquake forecasts (e.g., BEVINGTON and ROBINSON, 1992; GROSS and RUNDLE, 1998; KAGAN and JACKSON, 2000; SCHORLEMMER *et al.*, 2003). This test is used to evaluate the accuracy with which probability measure $P(\vec{x}_i, t_0, t_1, t_2)$ can forecast future ($t_2 < t < t_3$) large ($m \geq 5$) events, relative to forecast from the null hypothesis. The likelihood $L$ is a probability measure that can be used to assess the utility of one forecast measure over another. Typically, one computes the logarithm of the likelihood ($\log_{10} L$) for the proposed measure $L$ and compare that to the likelihood measure $L_N$ for a representative null hypothesis. The ratio of these two values then yields information about which measure is more accurate in forecasting future events.

In the likelihood test, a probability density function is required. Following HOLLIDAY *et al.* (2004a, b), we use a global Gaussian model and a local Poissonian model in our study. The use of the global Gaussian model was proposed for test the PI method forecast (TIAMPO *et al.*, 2002a). The second model used is based on work performed by the Regional Earthquake Likelihood Methods (RELM) group (SCHORLEMMER *et al.*, 2003). The likelihood values for the PI method forecast are defined as $L_G$ for the global Gaussian model and $L_P$ for the local Poissonian model. Similarly, the likelihoods for the null hypothesis are defined as $L_{GN}$ for the global Gaussian model and $L_{PN}$ for the local Poissonian model.

We first take the global Gaussian model. Then we compute the log-likelihood $\log_{10}(L_G)$ for the forecast of Figure 1 and the log-likelihood $\log_{10}(L_{GN})$ for the seismic intensity map in Figure 2. The results are summarized in the row of No. 4 in Table 1. The computed log-likelihood $\log_{10}(L_G) = -47.0$ is larger than the log-likelihood $\log_{10}(L_{GN}) = -58.9$. Next we take the local Poissonian model. As done for the global Gaussian model, the log-likelihoods are computed for the forecast of Figure 1, $\log_{10}(L_P) = -64.7$, and for the seismic intensity map in Figure 2, $\log_{10}(L_{PN}) = -108.7$ (Table 1). The value of $\log_{10}(L_P)$ is larger than that of $\log_{10}(L_{PN})$. Since larger values of the log-likelihoods indicate a more successful hypothesis, the logical conclusion is that the optimized PI method has better forecast skill than the actual seismic intensity data.

We also carry out statistical testing for the other regions (Nos. 1, 2, 3, and 5 in Table 1). We take $t_0$ = Jan. 1, 1960, $t_1$ = Jan. 1, 1968, $t_2$ = Dec. 31, 1994, and $t_3$ = Mar. 14, 2004 as done for the visual inspections. We first take the global Gaussian model. For all regions we find that the log-likelihood $\log_{10}(L_G)$ is larger than that $\log_{10}(L_{GN})$. We next take local Poissonian model. We again find that the log-likelihood $\log_{10}(L_P)$ is larger than that $\log_{10}(L_{PN})$ for all regions. These results support our conclusion that the method has forecasting skill.

*5. Forecasting large earthquakes for the period 2000-2009 in the Kobe and Tokyo regions*

TIAMPO *et al.* (2002b) and RUNDLE *et al.* (2002, 2003) proposed that the true test of any forecasting algorithm is to make a future forecast that proves correct. The diffusive, mean field nature of the dynamics (FISHER *et al.*, 1997; FERGUSON *et al.*, 1997; RUNDLE *et al.*, 2000b; KAGAN and JACKSON, 2000) leads to a hypothesis that forecasts should convey information for time $t$ approximately in the range: $t_3 \{= t_2 + (t_2 - t_1)\} > t > t_2$.

According to this hypothesis, previous studies (RUNDLE *et al.*, 2003; TIAMPO *et al.*, 2002a; HOLLIDAY *et al.*, 2004) took the length of the change interval to be equal to the length of the forecast interval for future large event forecast during the period of 2000-2009 in Southern California. Here we also do this for a future forecast in the period from 2000 to 2009 ($t_3 - t_2 = 10$ years) for the Tokyo and Kobe regions. For both regions, we assume that $t_3 - t_2 = t_2 - t_1 = 10$ years. That is, the times used are $t_1 =$ Jan. 1, 1989 and $t_2 =$ Dec. 31, 1999. We optimize the initial time $t_0$ for the forecasting large earthquakes in the period from Jan. 1, 2000 to Mar. 14, 2004 by using statistical testing.

*5.1 Kobe Region*

We first consider the Kobe region (No. 4 in Table 1) where there are 6 earthquakes having $m \geq 5$ occurring in the period 2000-2004. Statistical testing using local Gaussian and local Poissonian models indicates that the time $t_0$ is optimized to be $t_0 = 1980$ (years) for region 4. Our forecast map is shown in Figure 3. We use times $t_1 =$ Jan. 1, 1989 and $t_2$ = Dec. 31, 1999 with the optimized initial time $t_0$ = Jan. 1, 1980. Values of $\log_{10}\{P'(\bar{x}_i, t_0, t_1, t_2)/P'_{MAX}\}$ are given using the same color code as in Figure 1. Inverted triangles are events, $m \geq 5$, during 1989-1999. Events of $m \geq 5$ that occurred between Jan. 1 2000 to Mar. 14, 2004 are plotted with circles. The 2000 Tottori earthquake (133.4° East long., 35.3° North lat.) again falls into one of the forecasted areas. This figure calls attention to regions that seem to be at risk for larger earthquakes having $m \geq 5$ during 2000-2009. The regions most at risk (orange and red coloured) include the epicentral area of the 1995 Kobe earthquake approximately located at 135.0° east longitude and 34.5° north latitude, the Wakayama area approximately located at 135.1° east longitude and 34.0° north latitude, the Mie area approximately located at 136.1° east longitude and 34.3° north latitude, and the Aichi area approximately located at 136.5° east longitude and 35.2° north latitude.

## 5.2 Tokyo Region

Next the PI method is applied for forecasting large future earthquakes in the Tokyo region. Since the Sep. 1, 1923 Kanto earthquake ($m = 7.9$) that killed more than 140,000 people, the Tokyo region has spent a seismically quiet period (e.g., RIKITAKE, 1990). However, earthquakes that could inflict disaster on the Tokyo region were pointed out (e.g., MOGI, 1985). Moreover, using the exponential probability distribution model for recurrence times, FERRAES (2003) estimated that a damaging earthquake ($m \geq 6.4$) may occur before the year June 2009. In these circumstances, it will be interesting to make the PI hazard map for forecasting large ($m \geq 5$) future (2000-2009) earthquakes for the Tokyo region.

As was done for the Kobe region, we assume times $t_1$ = Jan. 1, 1989 and $t_2$ = Dec. 31, 1999 and utilize initial time $t_0$ that is optimized for the forecast of large ($m \geq 5$) large quakes in the period from Jan. 1, 2000 to the present (Mar. 14, 2004) by the statistical tests. The optimized value is $t_0$ = Jan. 1, 1965. The forecast map is shown in Figure 4. Values of $\log_{10}\{P'(\bar{x}_i, t_0, t_1, t_2)/P'_{MAX}\}$ are given using the color code as in Figure 2. Earthquakes with $m \geq 5$ in the change and forecast intervals are represented by inverted triangles and circles, respectively. The 2000 Miyake earthquake swarm falls in colour coded anomalies in Figure 4. For comparison, the spatial distribution of relative seismic intensities for region 6 for the time period 1965 to 1999 is given in Figure 5. The relative intensity is defined as the ratio $n(\bar{x}_i, t_0, t_2)/n_{MAX}$ where $t_0$ = Jan. 1, 1965 and $t_2$ = Dec. 31, 1999.

The forecast map in Figure 4 was first presented by one of the authors (JBR) at a lecture on Oct. 13, 2004 at Kyoto Univ., Japan (Organizer: Professor James Mori, Kyoto Univ.) and at the International Conference on Geodynamics held on Oct. 14-16, 2004 at the Univ. of Tokyo, Japan (Organizer: Professor Mitsuhiro Matsu'ura, University of Tokyo). After this appearance, the Oct. 23, 2004 Niigata earthquake of $m$ = 6.8 (hereinafter referred to as the 2004 Niigata earthquake) occurred. The epicenter was 138.7° east longitude and 37.3° north latitude and the depth was 16 km. Large aftershocks with $5.0 \leq m$ occurred around the hypocenter of this earthquake. Figure 6 is the same as Figure 4 but these events are added in Figure 6. Note that the 2004 Niigata earthquake and its aftershocks fall into a colour coded anomaly, this indicates that the method may have considerable promise for forecasting future large earthquakes in Japan. Future monitoring of the Tokyo and Kobe regions will test the accuracy and reliability of our PI studies.

## 6. *Conclusion*

We have utilized the pattern informatics (PI) technique to obtain the seismic hazard in the Kobe and Tokyo regions of Japan. First, we introduced the PI method. Next we briefly described the JMA catalog with the locations, magnitudes, and times of earthquakes in and around Japan. Earthquakes of magnitude equal to or larger than $m_c = 3$ with depths shallower than 20 km in the time period from 1955 to 2004 were selected. We applied the method for optimal forecasting of large ($m \geq 5$) earthquakes in Kobe region during the time period Jan. 1, 1995 – Mar. 14, 2004 ($t_3 - t_2 \sim 9$ years). This was best illustrated using a specific example. The created forecast map demonstrated that retrospective forecast is reasonably successful. We noted that the 1995 Kobe and 2000 Tottori earthquakes fall in anomalous areas. For comparison, the seismic intensity distribution (null hypothesis) was shown. We found that the optimized PI method narrows the possible locations of the occurrence of large future earthquakes. Statistical testing for the PI method forecast was carried out. For a null hypothesis, the actual seismic intensity data were used. We used maximum likelihood tests with the probability density functions, global Gaussian model and local Poissonian model. The results of the statistical tests showed that the log-likelihood for the PI forecast was larger than that for the null hypothesis. We concluded that the optimized PI method has better forecast skill than the actual seismic intensity data. For further test of the forecasting algorithm, we make a future forecast to prove correct. The PI method was applied for creating potential maps for large earthquake events with $m \geq 5$ for the forecast time period from Jan. 1, 2000 to Dec. 31, 2009 in the Kobe and Tokyo regions. Several forecast areas on the Kobe map were pointed out: the epicentral area of the 1995 Kobe earthquake, the Wakayama area, the Mie area, and the Aichi area. The Tokyo forecast map was obtained prior to the Oct. 23, 2004 Niigata earthquake ($m = 6.8$) and the principal aftershocks with $5.0 \leq m$ occurred. These earthquakes occurred in an anomalous area in the Tokyo hazard map.   This is an example that the PI technique has considerable promise as an intermediate-term earthquake forecasting in Japan.

*Acknowledgements*

DLT acknowledges the National Science Foundation (USA) under grant NSF ATM-03-27571.   JBR and JRH are supported by a grant from US Department of Energy, Office of Basic Energy Sciences to the University of California, Davis DE-FG03-95ER14499, and through additional funding from the National Aeronautics and Space Administration under grant through the Jet Propulsion Laboratory to the University of California, Davis.   KZN thanks W. Klein for valuable and helpful discussions, JSPS Research Fellowship for Young Scientists for financial support, JMA for providing Japanese earthquake data, and H. Tsuruoka for discussion of the JMA earthquake data.

**Table caption**

Table 1

*Shown are the study regions used for forecasting large earthquakes in the Kobe and Tokyo regions and the statistical results. The regions are numbered. 0.1° corresponds roughly to 11 km. The number of 0.1° × 0.1° × 20 km boxes is given in the column "N". The log-likelihoods computed using the global Gaussian model and local Poissonian model for the PI method forecast are given in the columns $\log_{10}(L_G)$ and $\log_{10}(L_P)$, respectively. Similarly, the log-likelihoods computed using the global Gaussian model and local Poissonian model for the seismic intensity data (null hypothesis) are given in the columns $\log_{10}(L_{GN})$ and $\log_{10}(L_{PN})$, respectively.*

**Figure captions**

Figure 1

The optimized PI method forecast for the Kobe region (No. 4 in Table 1) for the period 1995-2004. Relative probabilities $\log_{10}\{P'(\bar{x}_i, t_0, t_1, t_2)/P'_{MAX}\}$ are given using the color code. The times used are $t_0$ = Jan. 1, 1960, $t_1$ = Jan. 1, 1968, and $t_2$ = Dec. 31, 1994. Earthquakes with $5.0 \leq m$ that took place during 1968-1994 are shown as inverted triangles. Circles represent events with $5.0 \leq m$ during the time period 1995-2004.

Figure 2

Relative seismic intensities $\log_{10}\{n(\bar{x}_i, t_0, t_2)/n_{MAX}\}$ for the Kobe region (No. 4 in Table 1) for the period from $t_0$ = Jan. 1, 1960 to $t_2$ = Dec. 31, 1994 (triangles: earthquakes with $5.0 \leq m$ in 1968-1994; circles: earthquakes with $5.0 \leq m$ in 1995-2004).

35 (Nanjo et al., 2004)

Figure 3

Pattern informatics method forecast for the Kobe region (No. 4 in Table 1) during the period 2000-2009. Relative probabilities $\log_{10}\{P'(\bar{x}_i,t_0,t_1,t_2)/P'_{MAX}\}$ are given using color code. The times used are also $t_0$ = Jan. 1, 1980, $t_1$ = Jan. 1, 1990, and $t_2$ = Dec. 31, 1999. Circles are events that occurred during the time period from 2000 to 2004 for again $5.0 \leq m$. Inverted triangles are events that occurred from 1990 to 1999 for again $5.0 \leq m$.

Figure 4

Pattern informatics method forecast for the Tokyo region (No. 6 in Table 1) for the period 2000-2009. Relative probabilities $\log_{10}\{P'(\bar{x}_i,t_0,t_1,t_2)/P'_{MAX}\}$ are given using color code. The times used are $t_0$ = Jan. 1, 1965, $t_1$ = Jan. 1, 1990, and $t_2$ = Dec. 31, 1999. Inverted triangles and circles represent events with $5.0 \leq m$ that have occurred during the time periods 1990 – 1999 and 2000 – 2004, respectively.

Figure 5

Relative seismic intensities $\log_{10}\{n(\bar{x}_i,t_0,t_2)/n_{MAX}\}$ for the Tokyo region (No. 6 in Table 1) for $t_0$ = Jan. 1, 1965 to $t_2$ = Dec. 31, 1999. Earthquakes with $5.0 \leq m$ that took place during 1990-1999 are shown as inverted triangles. Circles represent events with $5.0 \leq m$ that have occurred during the period 2000-2004.

Figure 6

Recreated forecast map for the Tokyo region (No. 6 in Table 1) for the period 2000-2009. This figure is the same as Figure 5, but the 2004 Niigata earthquake ($m$ = 6.8) and its large aftershocks ($5.0 \leq m$) are added. See also the caption of Figure 5.

| Region | No. | North lat. (deg.) | East long. (deg.) | Depth (km) | $N$ | $\mathrm{Log}_{10}(L_G)$ | $\mathrm{Log}_{10}(L_P)$ | $\mathrm{Log}_{10}(L_{GN})$ | $\mathrm{Log}_{10}(L_{PN})$ |
|---|---|---|---|---|---|---|---|---|---|
| Kobe | 1 | 34.1-35.2 | 134.2-135.8 | 20 | 176 | -17.0 | -32.4 | -21.2 | -49.6 |
| | 2 | 33.7-35.6 | 133.8-136.2 | 20 | 456 | -27.2 | -40.0 | -31.7 | -71.5 |
| | 3 | 33.3-36.0 | 133.4-136.6 | 20 | 864 | -34.3 | -47.4 | -39.8 | -81.2 |
| | 4 | 32.9-36.4 | 133.0-137.0 | 20 | 1500 | -47.0 | -64.7 | -58.9 | -108.7 |
| | 5 | 32.5-36.8 | 132.6-137.4 | 20 | 2064 | -48.4 | -67.0 | -63.2 | -114.2 |
| Tokyo | 6 | 33.0-38.0 | 136.0-142.0 | 20 | 3000 | - | - | - | - |

Table 1 (Nanjo *et al.*, 2004)

Fig. 1 (Nanjo et al., 2004)

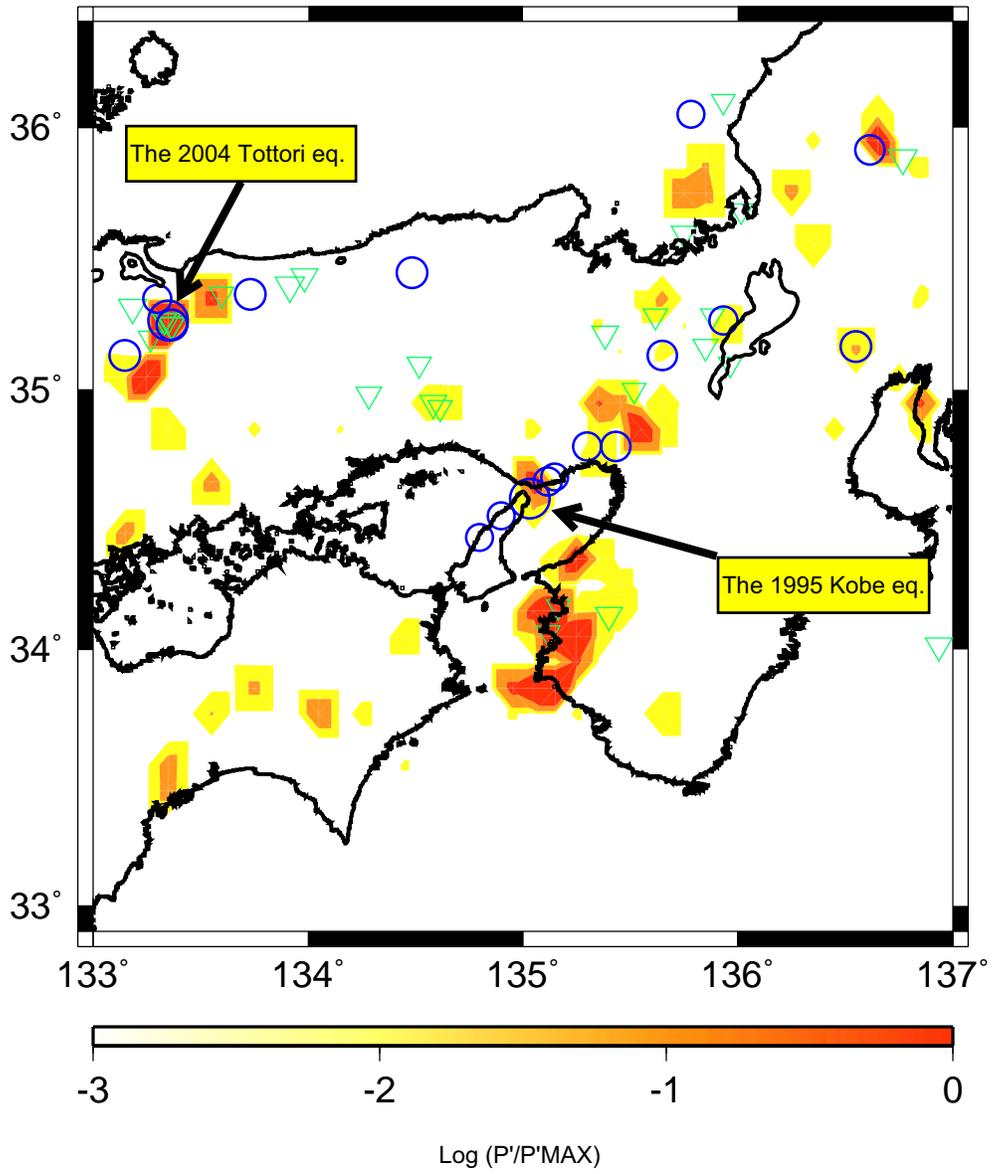

Fig. 2 (Nanjo et al., 2004)

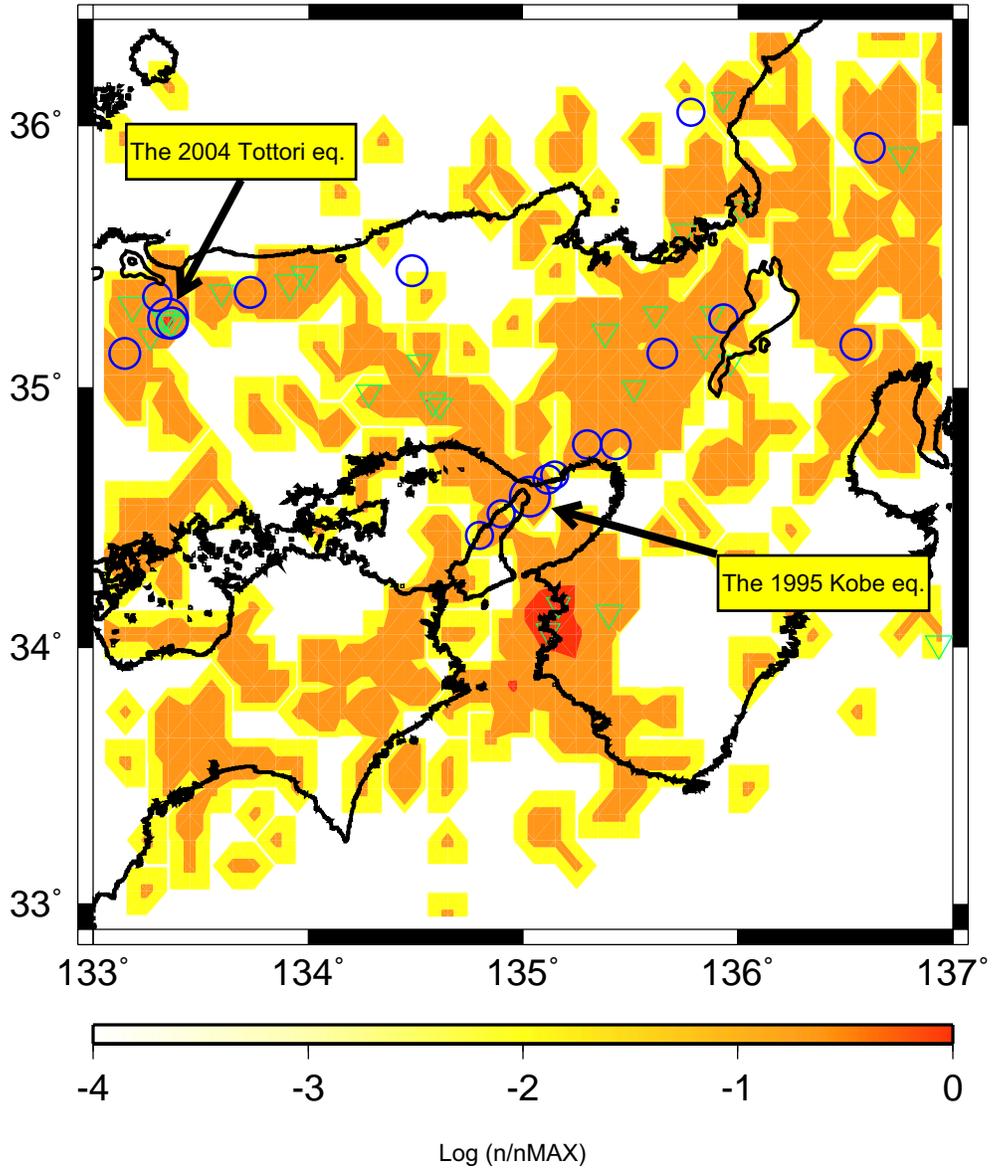

Fig. 3 (Nanjo et al., 2004)

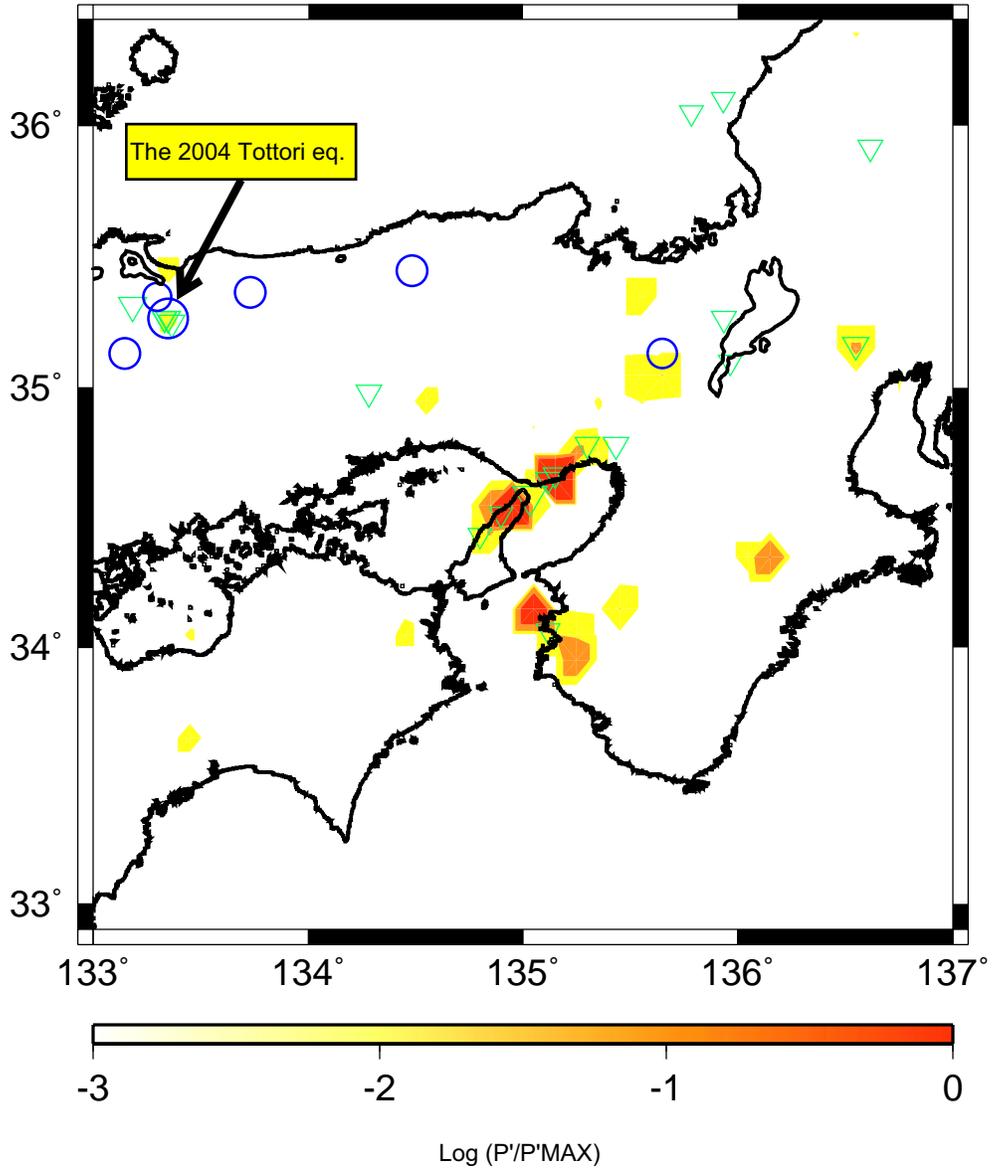

Fig. 4 (Nanjo et al., 2004)

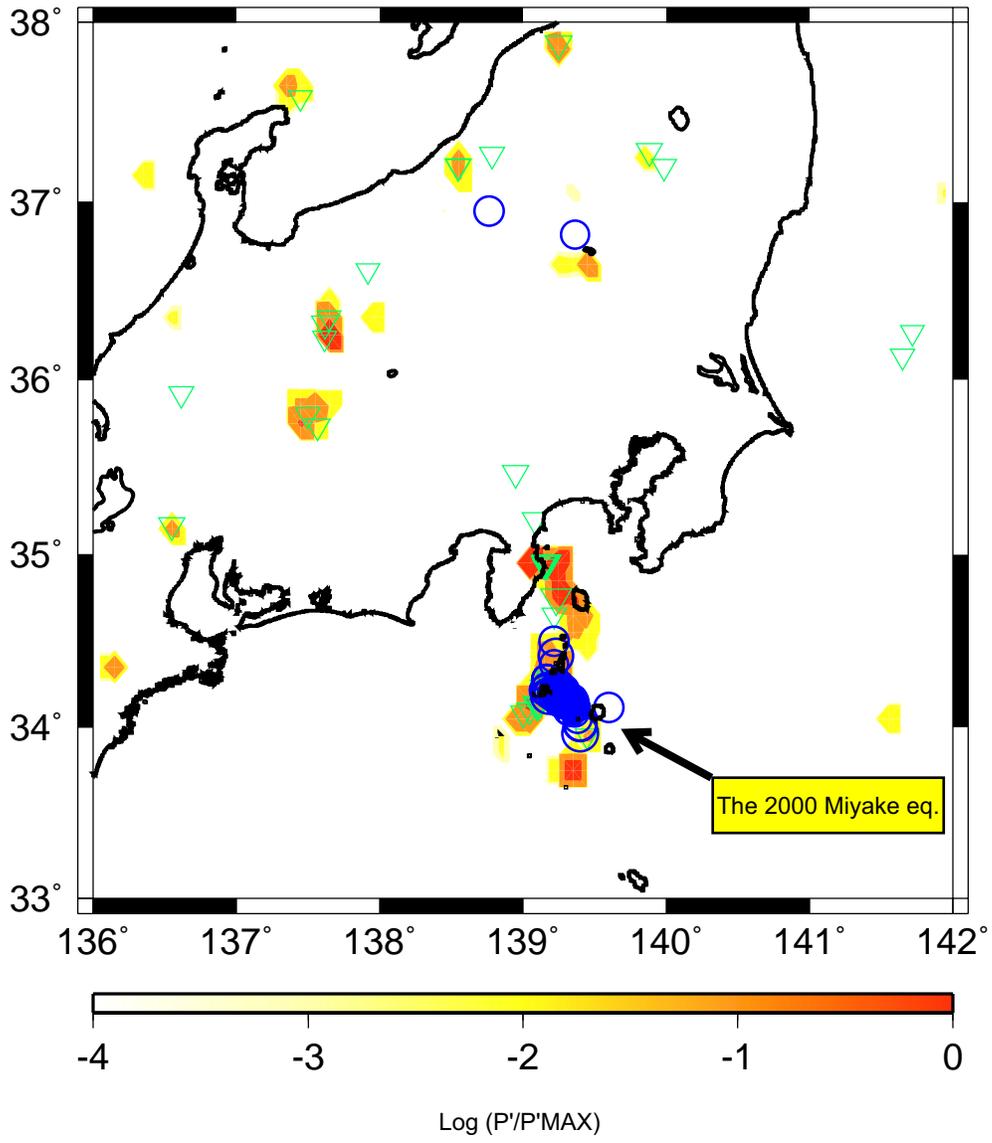

Fig. 5 (Nanjo et al., 2004)

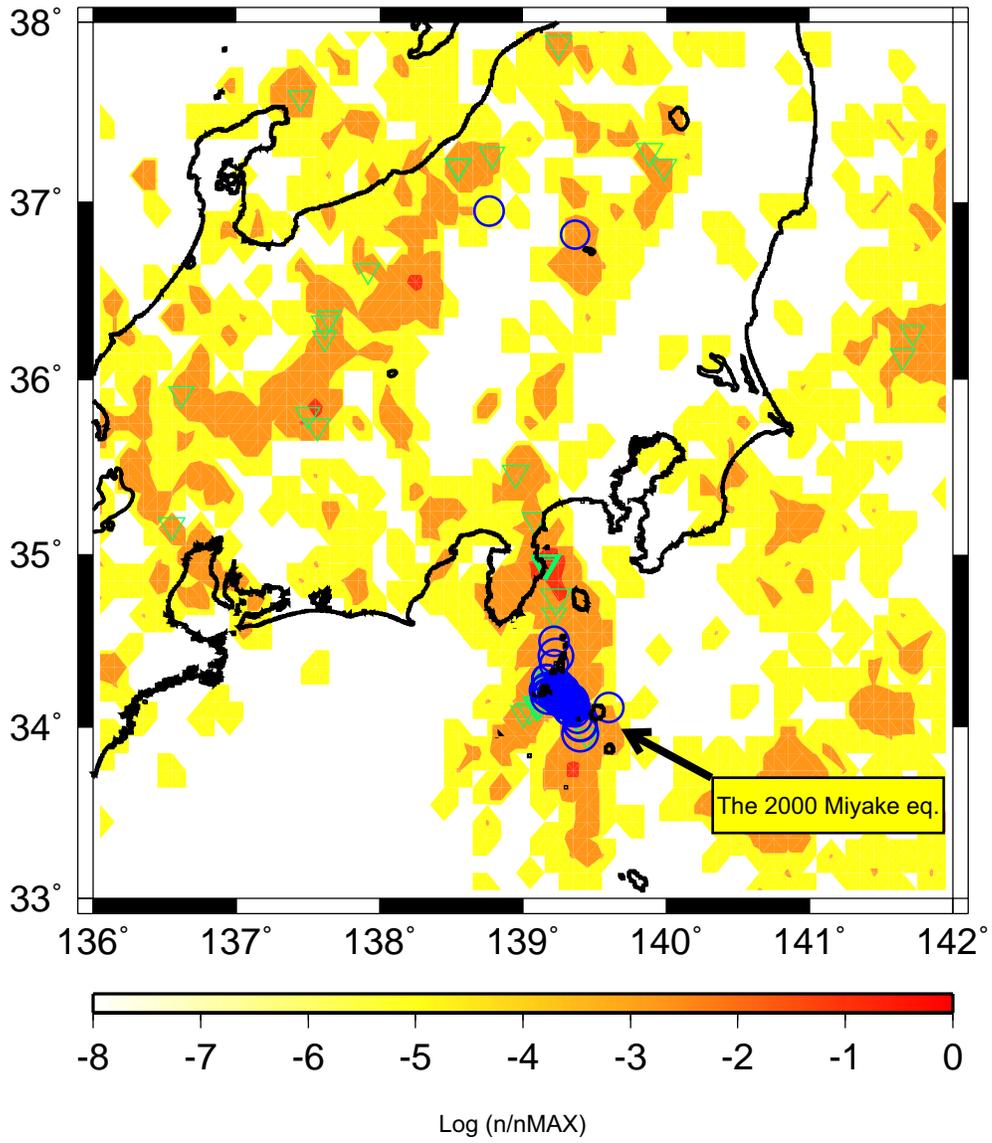

Fig. 6 (Nanjo et al., 2004)

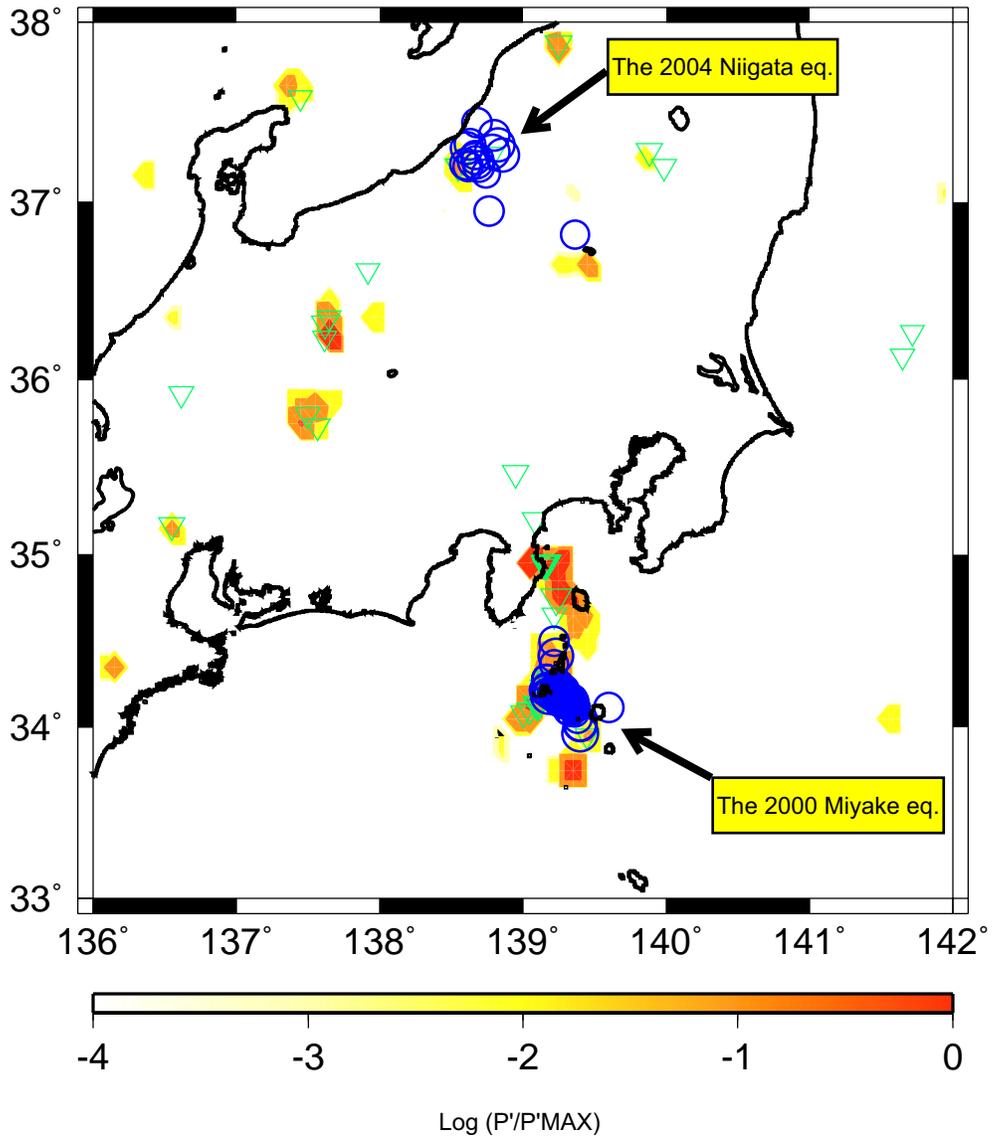